\documentclass[10pt,conference]{IEEEtran}
\IEEEoverridecommandlockouts
\usepackage{adjustbox}
\usepackage{algorithm}
\usepackage{algorithmic}
\usepackage{amsfonts}
\usepackage{amsmath}
\usepackage{amssymb}
\usepackage{array}
\usepackage{colortbl}
\usepackage{diagbox}
\usepackage{float}
\usepackage{graphicx}
\usepackage{hyperref}
\usepackage{inputenc}
\usepackage{lipsum}
\usepackage{makecell}
\usepackage{multirow}
\usepackage{pgf-pie}
\usepackage{pgfplotstable}
\usepackage{pgfplots}
\usepackage{slashbox}
\usepackage{tabularx}
\usepackage{textcomp}
\usepackage{tcolorbox}
\usepackage{tikz}
\usepackage{xcolor}
\usepackage{tikz}
\usepackage{textcomp}
\usetikzlibrary{shapes.geometric, arrows}

\tikzstyle{startstop} = [rectangle, rounded corners, minimum width=3cm, minimum height=1cm, text centered, draw=white, fill=gray!30]
\tikzstyle{process} = [rectangle, minimum width=3cm, minimum height=1cm, text centered, draw=white, fill=gray!20]
\tikzstyle{arrow} = [thick,->,>=stealth]
\usetikzlibrary{patterns, shapes.geometric, arrows, positioning, calc}
\pgfplotsset{compat=1.18}

\begin{document}

\author{\IEEEauthorblockN{ Seyed Moein Abtahi}
\IEEEauthorblockA{
\textit{Faculty of Engineering and Applied Science}\\
\textit{Ontario Tech University}\\
seyedmoein.abtahi@ontariotechu.net}
\and
\IEEEauthorblockN{ Akramul Azim}
\IEEEauthorblockA{
\textit{Faculty of Engineering and Applied Science}\\
\textit{Ontario Tech University}\\
akramul.azim@ontariotechu.ca}
}

\title{Securing LLM-Generated Embedded Firmware through AI Agent-Driven Validation and Patching}

\maketitle

\begin{abstract}
Large Language Models (LLMs) show promise in generating firmware for embedded systems but often introduce security flaws and fail to meet real-time performance constraints. This paper proposes a three-phase methodology that combines LLM-based firmware generation with automated security validation and iterative refinement in a virtualized environment. Using structured prompts, models like GPT-4 generate firmware for networking and control tasks, deployed on FreeRTOS via QEMU. These implementations are tested using fuzzing, static analysis, and runtime monitoring to detect vulnerabilities such as buffer overflows (CWE-120), race conditions (CWE-362), and denial-of-service threats (CWE-400). Specialized AI agents for Threat Detection, Performance Optimization, and Compliance Verification collaborate to improve detection and remediation. Identified issues are categorized using CWE, then used to prompt targeted LLM-generated patches in an iterative loop. Experiments show a 92.4\% Vulnerability Remediation Rate (37.3\% improvement), 95.8\% Threat Model Compliance, and 0.87 Security Coverage Index. Real-time metrics include 8.6ms worst-case execution time and 195\textmu s jitter. This process enhances firmware security and performance while contributing an open-source dataset for future research.
\end{abstract}


\section{Introduction}

Modern embedded and real-time control systems are critical in mission-critical applications ranging from automotive subsystems and aerospace controllers to industrial automation, where network connectivity facilitates functionalities such as over-the-air (OTA) updates, remote diagnostics, and real-time data exchange. However, this enhanced connectivity concurrently introduces novel attack vectors, thereby increasing the risk of exploitation in embedded devices. Large Language Models (LLMs), including GPT-3.5 \cite{OpenAI_GPT35}, GPT-4 \cite{OpenAI_GPT4}, and Code Llama \cite{touvron2023llama}, have demonstrated substantial promise in generating and debugging embedded software \cite{englhardt2024exploring}. Despite their syntactic accuracy, these models frequently lack the necessary domain-specific contextual understanding, resulting in semantically flawed code that may suffer from misinterpreted functions, API hallucinations, and omitting essential security checks. Such issues amplify the reliability and security challenges in embedded systems \cite{agarwal2024codemirage, zhang2024llm}.

AI agents offer task-specific capabilities that complement LLMs by collaboratively addressing security and performance in firmware pipelines \cite{deng2025ai, tran2025multi}. However, they also introduce risks such as prompt injection and task hijacking, especially in adversarial settings. Traditional fuzz testing techniques, including coverage-guided approaches like AFL++ \cite{fioraldi2020afl++} and Syzkaller \cite{mallissery2023demystify}, excel at identifying conventional software vulnerabilities; however, they are not specifically designed to detect anomalies typical of LLM-generated code, such as incorrect protocol state transitions, incomplete input validations, and faulty memory allocations \cite{chong2024artificial, paria2024spell}. In response, our research performs random input fuzzing to QEMU \cite{bellard2005qemu}, checking logs for anomalies such as deadline misses and ``overflow'' strings, thereby enabling the detection of LLM-specific vulnerabilities in embedded firmware. While security frameworks such as SecRT-LLM and SPELL have effectively leveraged LLMs for hardware vulnerability detection and secure SoC design \cite{paria2024spell,zhang2024ecg}, the integration of LLM-driven firmware generation with robust, software-based security validation in networked embedded environments remains largely unexplored. This study addresses this gap by proposing a comprehensive framework that unifies LLM-assisted code generation with advanced security validation techniques, ensuring that every stage, from code generation and fuzz testing to vulnerability detection and remediation, is meticulously aligned with both functional and real-time operational requirements.

\begin{table*}[t]
\centering
\setlength{\tabcolsep}{4pt}  
\caption{Summary of LLM Applications in Embedded Systems and Security}
\scalebox{1}{
\renewcommand{\arraystretch}{1.2}  
\begin{tabular}{|>{\centering\arraybackslash}p{2cm}|>{\centering\arraybackslash}p{2.5cm}|>{\centering\arraybackslash}p{2.5cm}|>{\centering\arraybackslash}p{3cm}|>{\centering\arraybackslash}p{2.5cm}|}
\hline
\rowcolor{gray!25}
\textbf{Paper} & \textbf{Focus Area} & \textbf{LLM Usage} & \textbf{Key Findings} & \textbf{Limitations} \\
\hline
\rowcolor{white}
Engelhardt et al. (2024) & Embedded Dev. \& Debug & LLM evaluation for hardware tasks & Aids in driver dev., I2C interfaces, power optimization & API hallucination, hardware spec confusion \\
\hline
\rowcolor{gray!10}
Dunne \& Fischmeister (2024) & Embedded Network Security & Fuzzing LLM network code & LLMs help with protocols but add vulnerabilities & No built-in verification \\
\hline
\rowcolor{white}
Paria et al. (2024) & Secure SoC Design & Code transformation \& security checks & Identifies/patches HW weaknesses with SecRT-LLM & Requires formal validation \\
\hline
\rowcolor{gray!10}
Zhang et al. (2024) & Embedded OS Fuzzing & Fuzzing corpora generation & 23\% coverage gain; 32 new kernel bugs & System call understanding dependent \\
\hline
\rowcolor{white}
Saha et al. (2024) & HW Security Benchmarking & Vulnerable HW database creation & 10,000 RTL designs with 16 security flaws & Manual vulnerability review needed \\
\hline
\end{tabular}}
\label{tab:llm_research}
\end{table*}

\section{Literature Review}

Recent studies have evaluated the potential and limitations of large language models (LLMs) in the context of embedded system development and security. Engelhardt et al. \cite{englhardt2024exploring} systematically evaluated GPT-3.5, GPT-4, and PaLM 2 to generate low-level embedded code, demonstrating that although iterative refinement enhances code correctness, significant security gaps persist in memory handling and peripheral interactions. Complementing these findings, Dunne and Fischmeister \cite{dunne2024weaknesses} conducted an extensive fuzzing study on LLM-generated networking stacks, revealing notable vulnerabilities in boundary checks and pointer management, especially in message parsing functions.

While existing frameworks like SecRT-LLM and SPELL \cite{paria2024spell, saha2024empowering} have primarily addressed hardware security risks, they do not adequately cover the unique challenges presented by LLM-generated firmware. This gap requires an approach that combines software security validation with automated LLM-based remediation. In this context, Zhang et al. \cite{zhang2024ecg} introduced ECG, an LLM-assisted technique that augments embedded operating system fuzzing by automating input specification creation and payload generation. Their approach significantly improves test coverage and uncovers previously undetected bugs, demonstrating the efficacy of LLMs in enhancing fuzzing workflows for embedded kernels.

Further exploring the role of LLMs in embedded development, Engelhardt et al. \cite{englhardt2024exploring} provide a comprehensive evaluation of LLMs for hardware interfacing tasks, highlighting both the potential for rapid prototyping and the challenges of API hallucinations and hardware specification misunderstandings. Their work underscores the importance of iterative refinement and domain-specific prompt engineering to mitigate these issues. In parallel, Dunne and Fischmeister \cite{dunne2024weaknesses} emphasize that while LLMs can scaffold protocol-handling code effectively, they frequently omit critical security validations, such as proper boundary checking and pointer arithmetic, thus necessitating robust post-generation verification processes like fuzz testing and static analysis.

Paria et al. \cite{paria2024spell} further contribute to the field by presenting an end-to-end tool flow for secure system-on-chip (SoC) design using LLMs. Their methodology leverages LLM-driven code generation alongside formal verification tools to detect and remedy complex hardware vulnerabilities, including dynamic deadlocks and finite-state machine (FSM) security violations. This multi-stage approach illustrates how iterative correction, guided by specialized prompts, can significantly enhance the security of hardware designs. Similarly, Saha et al. \cite{saha2024empowering} propose a pipeline for developing a vulnerability database for hardware security, which systematically introduces, detects, and labels known vulnerabilities, thus providing a valuable resource for both machine learning-based methods and traditional verification tools.

Collectively, these studies reveal that while LLMs offer promising capabilities in generating and refining embedded system code, they consistently require supplementary verification mechanisms to ensure reliability and security. The integration of LLM-based code generation with automated fuzz testing and formal security validations emerges as a critical research direction, particularly for addressing the challenges inherent in embedded networking and firmware applications. As summarized in Table \ref{tab:llm_research}, each work highlights the potential of LLMs while also identifying specific areas, such as API hallucination, incomplete domain knowledge, and the need for built-in security checks, that must be addressed to harness their capabilities in mission-critical applications fully.

\subsection*{Research Questions}

Based on the challenges and advancements discussed in the literature, the following research questions are central to the ongoing exploration of LLMs in embedded systems and hardware security:

RQ1: How can LLMs be effectively integrated into the embedded systems development and debugging lifecycle to improve code generation, debugging efficiency, and overall system performance?

RQ2: To what extent can LLM-based tools, such as SecRT-LLM, enhance the detection and correction of hardware vulnerabilities in SoC designs, and how do they compare to traditional methods?

RQ3: What are the limitations and challenges of using LLMs in the context of embedded systems and hardware security, and how can these limitations be mitigated through prompt engineering and verification methods?

\section{Methodology}
Recent research on LLM-generated firmware has revealed a pattern of recurring security flaws, including memory mismanagement, improper protocol implementation, and incorrect peripheral interactions. Studies evaluating LLM-generated code, such as those by Engelhardt et al., highlight that while these models can produce syntactically correct code, they often fail to meet security and timing constraints in embedded systems. Additionally, existing fuzz testing methodologies, including coverage-guided fuzzing tools like AFL++ and Syzkaller, have demonstrated effectiveness in detecting software vulnerabilities but are not specifically optimized for the unique characteristics of LLM-generated firmware.

Given these findings, our approach builds on existing methodologies by introducing a structured, iterative pipeline that integrates fuzz testing, static analysis, and automated remediation through LLM-driven refinement. This methodology is designed to systematically identify and mitigate vulnerabilities in LLM-generated firmware by leveraging software-based testing frameworks in a controlled virtualized environment.

\subsection{Three-Phase Generation Process}
The proposed methodology follows an iterative, three-phase process that combines LLM-driven firmware generation with rigorous security validation in a virtualized real-time environment. In the initial phase, we extract and formalize protocol specifications from existing documentation, which are then used to prompt Large Language Models (LLMs) such as GPT models or Llama models to generate baseline firmware. An automated validation process ensures syntax correctness and adherence to fundamental functional requirements. In the second phase, the generated firmware is deployed in a virtualized real-time setup where comprehensive security testing is performed. This includes intelligent fuzz testing targeting protocol edge cases and boundary conditions, complemented by static analysis to detect unsafe memory operations and buffer overflows. 

Runtime monitoring further validates that the firmware meets stringent real-time performance constraints. Finally, the third phase focuses on iterative patch refinement. Detected vulnerabilities are documented and fed back into an automated LLM-based remediation pipeline. Structured prompts guide the LLM to generate targeted security patches, which are then re-tested in the virtualized environment. This entire process is augmented by specialized AI agents that work alongside the LLM, providing domain-specific expertise in threat detection, performance optimization, and compliance verification. Quantitative metrics such as worst-case execution time and task jitter are used to measure improvements, ensuring continuous enhancement of both functionality and security through this collaborative human-AI framework.

\subsection{Threat Model}

In the context of LLM-generated firmware for embedded systems, we identify three primary security threats that our methodology aims to address: buffer overflow, race conditions, and denial of service (DoS) \cite{gu2007denial}. The buffer overflow occurs when data is written beyond the allocated memory buffer, potentially leading to arbitrary code execution or system crashes; LLMs may generate code that lacks proper bounds checking, making this a critical concern. Race conditions arise when multiple tasks access shared resources without adequate synchronization, resulting in data corruption or inconsistent states, which is particularly risky in the concurrent environments common to embedded applications. DoS threats involve overwhelming the system with excessive inputs or requests, causing it to miss real-time deadlines or become unresponsive, especially since LLMs might not inherently optimize for resource constraints. Our methodology incorporates targeted testing strategies, including fuzz testing, static analysis, and stress testing, to detect and mitigate these threats, as detailed in the subsequent sections.

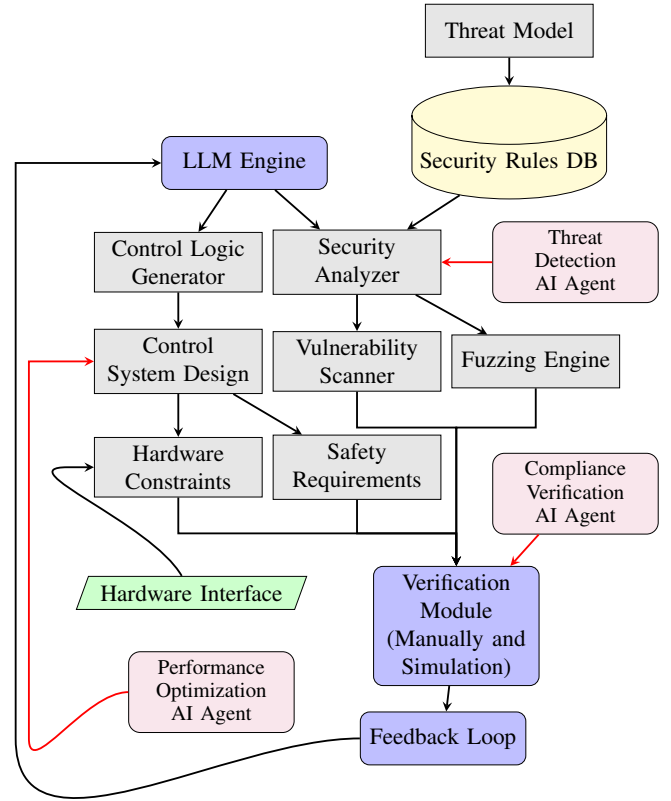
\begin{figure}[ht]
    \scalebox{0.88}{
    \centering
    \begin{tikzpicture}[
        node distance=1.5cm,
        scale=1,
        every node/.style={scale=1},
        start/.style={
            rectangle,
            rounded corners,
            minimum width=2.5cm,
            minimum height=0.8cm,
            text centered,
            draw=black,
            fill=blue!25
        },
        process/.style={
            rectangle,
            minimum width=2.3cm,
            minimum height=0.8cm,
            text centered,
            text width=2.3cm,
            draw=black,
            fill=gray!20
        },
        database/.style={
            cylinder,
            shape border rotate=90,
            aspect=0.3,
            minimum height=0.8cm,
            minimum width=1.8cm,
            draw=black,
            fill=yellow!20
        },
        interface/.style={
            trapezium,
            trapezium left angle=70,
            trapezium right angle=110,
            minimum width=1.0cm,
            minimum height=0.5cm,
            text centered,
            draw=black,
            fill=green!20
        },
        agent/.style={
        rectangle,
        rounded corners=5pt,
        draw=black,
        fill=purple!10,
        minimum width=2.5cm,
        minimum height=1cm,
        text centered,
        align=center,
        font=\small
        },
        arrow/.style={
        thick,
        ->,
        >=stealth
        }
        ]
        \node (llm) [start] at (-4.5,-0.5) {LLM Engine};
        
        \node (threat_agent) [agent] at (0.5,-2) {Threat\\ Detection\\ AI Agent};
        \node (perf_agent) [agent] at (-5,-8.5) {Performance\\ Optimization\\ AI Agent};
        \node (compliance_agent) [agent] at (0.5,-5.5) {Compliance\\ Verification\\ AI Agent};
        
        \node (clg) [process] at (-5.5,-2) {Control Logic Generator};
        
        \node (sa) [process] at (-2.8,-2) {Security Analyzer};
        
        \node (csd) [process] at (-5.5,-3.5) {Control System Design};
        
        \node (vs) [process] at (-2.8,-3.5) {Vulnerability Scanner};
        \node (fe) [process] at (-0.1,-3.5) {Fuzzing Engine};
        
        \node (hc) [process] at (-5.5,-5.1) {Hardware Constraints};
        \node (sr) [process] at (-2.8,-5.1) {Safety Requirements};
        
        \node (hi) [interface] at (-5.3,-7) {Hardware Interface};
        \node (db) [database] at (-0.5,-0.5) {Security Rules DB};
        
        \node (tm) [process] at (-0.5,1.5) {Threat Model};
        
        \node (vm) [start, align=center] at (-1.3,-7.5) {Verification\\ Module \\ (Manually and\\ Simulation)};
        \node (fl) [start] at (-1.5,-9.2) {Feedback Loop};
        
        \draw [arrow] (llm) -- (clg);
        \draw [arrow] (llm) -- (sa);
        
        \draw [arrow] (clg) -- (csd);
        \draw [arrow] (csd) -- (hc);
        \draw [arrow] (csd) -- (sr);
        
        \draw [arrow] (sa) -- (vs);
        \draw [arrow] (sa) -- (fe);
        \draw [arrow] (db) -- (sa);
        
        \draw [arrow,draw=red] (threat_agent) to[out=180,in=0] (sa);
        \draw [arrow,draw=red] (perf_agent) to[out=180,in=270] ([xshift=-5cm]fl.west) |- (csd);
        \draw [arrow,draw=red] (compliance_agent) -- (vm);
        
        \draw [arrow] (tm) -- (db);
        
        \draw [arrow] (hi) to[out=120,in=180] ([xshift=-0.4cm]hc.west) |- (hc);        
        \draw [arrow] (hc) -- ++(0,-1) -| (vm);
        \draw [arrow] (sr) -- ++(0,-1) -| (vm);
        \draw [arrow] (vs) -- ++(0,-1) -| (vm);
        \draw [arrow] (fe) -- ++(0,-1) -| (vm);
        
        \draw [arrow] (vm) -- (fl);
        \draw [arrow] (fl) to[out=180,in=270] ([xshift=-5.2cm]fl.west) |- (llm);
    \end{tikzpicture}}
    \caption{LLM-based Control System Architecture with AI Agent Integration and Security Modeling}
    \label{fig:architecture}
\end{figure}

The threat model plays a pivotal role in our iterative pipeline by informing the security rules and validation processes that ensure the robustness of LLM-generated firmware. As illustrated in Figure \ref{fig:architecture}, the Threat Model component directly influences the Security Rules Database, which in turn supports the Security Analyzer in identifying vulnerabilities such as buffer overflows, race conditions, and denial of service attacks. By systematically incorporating these threats into our testing and remediation cycles, we enhance the security and reliability of the firmware, ensuring it meets the stringent requirements of embedded systems.

\subsection{LLM-Based Control System Architecture with Security Integration}
To ensure a comprehensive security validation framework, our methodology follows a structured LLM-based control system architecture, as illustrated in Fig. \ref{fig:architecture}. This architecture integrates multiple components, including a control logic generator, security analyzer, vulnerability scanner, fuzzing engine, verification module, and specialized AI agents, which collectively facilitate the iterative refinement of LLM-generated firmware.
The architecture consists of an LLM engine responsible for generating initial firmware code, which is then processed by the control logic generator to refine control system design aspects. 

The security analyzer identifies potential vulnerabilities using a combination of static analysis and runtime security validation, aided by a vulnerability scanner, fuzzing engine, and a dedicated Threat Detection Agent. The architecture leverages a security rules database to validate firmware compliance with predefined security constraints to enhance security. The Performance Optimization Agent analyzes runtime characteristics and suggests improvements to the control system design, while the Compliance Verification Agent ensures adherence to industry standards during verification. The verification module systematically evaluates firmware integrity and provides feedback for iterative improvement, ensuring security and real-time performance adherence.

\subsection{Virtualized Real-Time Setup}

To accurately evaluate the performance and security of LLM-generated firmware, we deploy a real-time validation framework within a virtualized environment. This framework leverages a lightweight real-time operating system (RTOS) \cite{hambarde2014survey,yun2022fuzzing}, such as FreeRTOS, running on an emulator like QEMU or within a container-based RT kernel simulation. The virtualized environment is configured to replicate key real-time characteristics, including task scheduling priorities, interrupt latencies, and deadline adherence.

By adopting this software-based approach, we conduct rigorous security and reliability testing without requiring specialized embedded hardware. The virtual RTOS setup enables efficient fault injection, performance analysis, and iterative refinement, ensuring that LLM-generated firmware meets the timing and safety constraints required for embedded applications.

\subsection{LLM-Driven Firmware Generation}

Using structured prompt engineering techniques, we instruct an LLM, such as GPT-4, to generate firmware modules tailored for specific embedded tasks. The generated code includes implementations for networking protocols, such as MQTT \cite{dinculeanua2019vulnerabilities} (with only minimal detection and parsing), as well as real-time control tasks, including sensor data acquisition and event-driven processing. Emphasis is placed on memory safety, error handling, and protocol adherence.

Rather than solely relying on LLM-generated firmware as a final product, our methodology treats these outputs as an initial draft subject to iterative refinement. Through controlled testing, we intentionally expose weaknesses in the generated code, feeding discovered issues back into the LLM to produce improved iterations with enhanced security and stability.

\subsection{Security and Reliability Testing}

The generated firmware undergoes comprehensive security and reliability testing through both fuzz testing and static analysis. The fuzz testing process involves injecting malformed or random data packets into the virtualized firmware environment to detect common vulnerabilities, such as buffer overflows, unintended state transitions, and out-of-bounds memory accesses. Tools such as AFL++ and Syzkaller perform coverage-guided fuzz testing, maximizing the exposure of security flaws within the firmware.

Static analysis tools, including Clang Static Analyzer and Cppcheck, identify unsafe pointer operations, race conditions, and violations of real-time constraints. Additionally, real-time performance validation measures worst-case execution time (WCET), task jitter, and deadline misses within the virtualized RTOS environment.

To ensure reproducibility, generated unit tests were executed with fixed random seeds. This guarantees deterministic outcomes across runs, preventing stochastic variations from impacting pass/fail counts. For exploratory testing, we also ran additional randomized test generations; however, reported metrics are based exclusively on deterministic configurations to ensure consistency.

\subsection{Iterative Vulnerability Remediation}
Security vulnerabilities and real-time violations are systematically documented and classified using Common Weakness Enumeration (CWE) references. The system integrates identified issues into a feedback loop. The Large Language Model (LLM) processes detailed error reports containing stack traces, memory dumps, and timing violations to generate potential remediation patches. This process is enhanced by specialized AI agents, with the Threat Detection Agent providing deeper context for identified vulnerabilities and the Performance Optimization Agent suggesting specific improvements to maintain real-time performance while implementing security fixes. This collaborative multi-agent approach requires minimal human validation before patches are merged into the codebase. Verification could occur through both direct system testing and QEMU emulation to confirm the complete resolution of vulnerabilities, ensuring system integrity and security compliance.

\begin{figure}[ht]
    \centering
    \scalebox{0.8}{
    \begin{tikzpicture}[node distance=1.5cm, auto, 
                        block/.style={rectangle, draw, fill=blue!20, 
                        text width=6cm, text centered, rounded corners, minimum height=1cm}]
        
        \node [block] (threat) {1. Threat Model \& Requirements};
        \node [block, below of=threat] (generate) {2. Generate / Design Code with LLM};
        \node [block, below of=generate] (integrate) {3. Integrate \& Build (FreeRTOS + QEMU)};
        \node [block, below of=integrate] (security) {4. Security \& Reliability Testing};
        \node [block, below of=security] (analyze) {5. Analyze Results};
        \node [block, below of=analyze] (refine) {6. Refine Code (LLM or Manual)};
        
        \draw [->] (threat) -- (generate);
        \draw [->] (generate) -- (integrate);
        \draw [->] (integrate) -- (security);
        \draw [->] (security) -- (analyze);
        \draw [->] (analyze) -- (refine);
        \draw [->] (refine) to [out=180,in=180] (integrate);
        
    \end{tikzpicture}
    }
    \caption{Firmware Development Cycle Integrating LLM-based Automation and Security Testing}
    \label{fig:methodology_flowchart}
\end{figure}
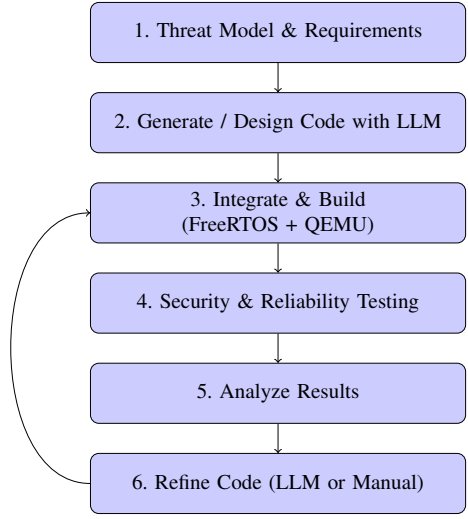

The LLM is prompted to generate a revised version of the firmware with targeted fixes for the reported issues, with improvements further refined by the Compliance Verification Agent to ensure industry standard adherence. This refined firmware is reintegrated into the testing framework, undergoing repeated validation cycles until a stable, secure, and real-time-compliant firmware iteration is achieved. By automating this iterative fuzz-patch cycle with intelligent agent oversight, the framework continuously improves LLM-generated firmware with minimal human intervention.
Figure \ref{fig:methodology_flowchart} illustrates the sequential development process, delineating each phase from threat modeling and initial code generation through to integration, testing, analysis, and refinement, with AI agents providing specialized expertise throughout the pipeline.

\subsection{AI Agent Integration}
To further enhance the robustness and adaptability of the LLM-driven firmware generation pipeline, a multi-agent AI ( Artificial Intelligence) framework is introduced to complement the existing methodology. These AI agents are strategically deployed across different stages of the development process, each assigned a specialization in firmware security, performance optimization, or vulnerability detection. Included among the specialized agents are a ``Threat Detection Agent", by which potential security vulnerabilities are proactively identified; a ``Performance Optimization Agent", through which runtime characteristics are analyzed and code improvements are suggested; and a ``Compliance Verification Agent", by which adherence to embedded system constraints and industry standards is ensured. In contrast to traditional single-model approaches, the multi-agent system allows for parallel analysis, cross-validation of findings, and more comprehensive code assessments to be conducted. Each agent is powered by specialized machine learning models that have been trained on domain-specific datasets, including embedded systems security databases, real-time operating system (RTOS) performance logs, and historical firmware vulnerability records. By integrating these AI agents, an additional layer of intelligent monitoring and optimization is provided, one that extends beyond static and dynamic analysis and results in a more adaptive and resilient firmware development framework.

\subsection{Evaluation Metrics}

 Multidimensional metrics are proposed to systematically assess the effectiveness of the LLM-based firmware generation pipeline and quantify the impact of AI agent integration. Both security and performance aspects are evaluated across development iterations. An objective comparison is enabled between baseline LLM-generated firmware and versions refined through the iterative, agent-augmented process.

\subsubsection{Security Assessment Metrics}
\begin{itemize}
  \item \textbf{Vulnerability Remediation Rate (VRR):} The percentage of vulnerabilities successfully remediated after each iteration cycle is calculated as:
    \begin{equation}
      \label{eq:vrr}
      VRR = \frac{\text{Vulnerabilities fixed}}{\text{Total vulnerabilities identified}} \times 100\%
    \end{equation}

  \item \textbf{Security Coverage Index (SCI):} A weighted metric quantifying the proportion of code paths examined during security testing is defined as:
  \begin{equation}
  \label{eq:SCI}
    SCI = \frac{w_1C_f + w_2C_s + w_3C_d}{w_1 + w_2 + w_3}
  \end{equation}
  where $C_f$ denotes fuzzing coverage, $C_s$ static analysis coverage, $C_d$ dynamic analysis coverage, and $w_i$ the corresponding weights.

  \item \textbf{Threat Model Compliance Score (TMCS):} The percentage indicating how effectively the firmware addresses the identified threats is given by:
  \begin{equation}
  \label{eq:TMCS}
    TMCS = \frac{\text{Mitigated threats}}{\text{Total threats identified}} \times 100\%
  \end{equation}
\end{itemize}

\subsubsection{Real-Time Performance Metrics}
\begin{itemize}
  \item \textbf{Worst-Case Execution Time (WCET):} The maximum time required for critical task completion is defined as:
  \begin{equation}
  \label{eq:WCET}
    WCET = \max_{i=1}^{n} {t_i}
  \end{equation}
  where $t_i$ represents the execution time for the $i$-th run.

  \item \textbf{Task Jitter (TJ):} The variance in task execution times is calculated as:
  \begin{equation}
  \label{eq:TJ}
    TJ = \sqrt{\frac{1}{n}\sum_{i=1}^{n}(t_i - \bar{t})^2}
  \end{equation}
  where $\bar{t}$ is the mean execution time.
\end{itemize}

\subsubsection{AI Agent Contribution Metrics}
\begin{itemize}
  \item \textbf{Agent Detection Accuracy (ADA):} For the Threat Detection Agent, the accuracy of vulnerability identification compared to ground truth is measured by:
  \begin{equation}
  \label{eq:ADA}
    ADA = \frac{TP + TN}{TP + TN + FP + FN}
  \end{equation}
  where $TP$, $TN$, $FP$, and $FN$ denote true positives, true negatives, false positives, and false negatives, respectively.
\end{itemize}

\subsubsection{Pipeline Efficiency Metrics}
\begin{itemize}
  \item \textbf{Iteration Efficiency Index (IEI):} The relative improvement per iteration cycle is measured as:
  \begin{equation}
  \label{eq:IEI}
    IEI = \frac{\Delta \text{Security Score} + \Delta \text{Performance Score}}{\text{Resources consumed}}
  \end{equation}
\end{itemize}

Metrics are collected at each iteration of the firmware development cycle. Baseline measurements are established using firmware generated solely by the LLM without agent augmentation, while subsequent iterations incorporate various combinations of AI agent contributions. Both the overall improvement of the methodology and the specific impact of individual agents and their collaborative interactions are thereby quantified. The metrics are visualized through radar charts to represent the multi-dimensional performance profile of each firmware version, facilitating comprehensive comparison across iterations. Through this systematic evaluation, the effectiveness of the approach in producing secure, reliable, and real-time-compliant embedded firmware is objectively demonstrated.

\section{Experimental Plan}
This research establishes a comprehensive methodology for enhancing the security posture and real-time performance of LLM-generated firmware within embedded systems environments through multi-agent AI integration. We employ OpenAI's GPT-4 in conjunction with specialized AI agents to generate protocol-specific message handlers through systematically structured prompt engineering, thereby maximizing protocol compliance while minimizing implementation vulnerabilities. Our approach implements a three-phase development and verification cycle enhanced by collaborative AI agent oversight.

Contemporary literature demonstrates the increasing adoption of LLMs within embedded systems and hardware security domains. Engelhardt et al. \cite{englhardt2024exploring} illustrate LLMs' efficacy as coding assistants when provided with appropriate hardware context, while Dunne et al. \cite{dunne2024weaknesses} identify critical vulnerabilities potentially introduced by unverified LLM-generated networking code. Further, frameworks developed by Paria et al. \cite{paria2024spell}, Saha et al. \cite{saha2024empowering}, and Zhang et al. \cite{zhang2024ecg} demonstrate effective methodologies for systematic verification and refinement of LLM-generated implementations. These findings inform our methodology, particularly addressing the challenges of integrating high-level design specifications with hardware constraints while mitigating domain-specific security vulnerabilities through multi-agent collaboration.

Our experimental implementation adheres to the three-phase iterative cycle with AI agent integration as presented in Fig. \ref{fig:methodology_flowchart}:

\subsection{Phase 1: Three-Phase Generation Process}
\begin{enumerate}
\item {Protocol Specification Extraction \& Formalization:} We extract and formalize protocol specifications from existing documentation, establishing comprehensive threat models addressing buffer overflow vulnerabilities (CWE-120), race conditions (CWE-362), and denial-of-service attack vectors (CWE-400) within documentation maintained in \texttt{main.c}.

\item{LLM-Driven Baseline Firmware Generation:} Utilizing GPT-4 with structured prompt engineering, we generate initial firmware implementations for embedded networking protocols and real-time control tasks through our custom \texttt{generate\_freertos\_task.py} and \texttt{llm\_refine.py} scripts, incorporating security parameters derived from threat modeling.

\item {Automated Validation \& Syntax Verification:} Generated components undergo automated validation for syntax correctness and adherence to fundamental functional requirements before integration into cohesive firmware images via \texttt{build\_and\_run.py} and project Makefiles.
\end{enumerate}

\begin{table*}[ht]
\centering
\setlength{\tabcolsep}{4pt}  
\caption{Comparison of LLM Applications in Embedded Systems and Security}
\scalebox{.8}{
\renewcommand{\arraystretch}{1.2}  
\begin{tabular}{|>{\centering\arraybackslash}p{2.5cm}|>{\centering\arraybackslash}p{2.8cm}|>{\centering\arraybackslash}p{2.8cm}|>{\centering\arraybackslash}p{2.8cm}|>{\centering\arraybackslash}p{2.8cm}|>{\centering\arraybackslash}p{2.1cm}|}
\hline
\rowcolor{gray!25}
\textbf{Approach / Paper} & \textbf{LLM Usage \& Domain} & \textbf{Fuzz Coverage / Specialized Testing} & \textbf{Concurrency / Real-Time Checks} & \textbf{Automated Patching} & \textbf{Public Dataset} \\
\hline
\rowcolor{white}
Engelhardt et al. (2024) & Evaluates GPT-based code generation for \textbf{embedded dev \& debugging}; multi-step tasks (I2C drivers, power mgmt). & Basic fuzz, mostly functional tests. Primarily focuses on \textbf{debugging} vs. deep security fuzz. & Minimal coverage of concurrency or real-time deadlines. Focus on driver correctness. & Iterative user feedback in a semi-manual loop (no fully automated patch pipeline). & No dedicated public dataset; main focus on measuring LLM success/failure in hardware tasks. \\
\hline
\rowcolor{gray!10}
Dunne \& Fischmeister (2024) & Investigates LLM auto-gen code for \textbf{network protocols} (PPP, AT commands). Identifies common boundary-check \& pointer issues. & Specialized protocol fuzz testing. Identified LLM "hallucinations" in pointer arithmetic and partial state checks. & Not a major focus; concurrency \& RT constraints not heavily tested beyond boundary checks. & Patches not directly integrated. Suggest manual "post-generation verification." & No open dataset; logs or fuzz inputs not systematically released. \\
\hline
\rowcolor{white}
Paria et al. (2024) & \textbf{SoC design} security (SecRT-LLM). LLM transforms/patches hardware RTL for secure compliance. & Custom coverage checks, formal verification steps for hardware design rules. & Real-time constraints not deeply addressed at the SoC level (focus on hardware security rules). & LLM-based iterative fix cycle for SoC designs—\textbf{semi-automated} flow, guided by linting/formal checks. & No standard public dataset of firmware flaws (centered on hardware design references). \\
\hline
\rowcolor{gray!10}
Zhang et al. (2024) & \textbf{ECG}: LLM-based fuzz corpus for \textbf{embedded OS} kernels, augmenting coverage. Found 32 new kernel bugs. & Gains up to 23\% coverage, synergy w/ Syzkaller or coverage-based fuzzers. Highly specialized for OS syscalls coverage. & Sched/RT constraints partially tested if kernel code includes scheduling modules, but not a main focus. & Uses a multi-stage LLM approach to refine fuzz inputs, not a direct patching approach. & The new kernel bugs discovered are not systematically organized into a public dataset. \\
\hline
\rowcolor{white}
Our Project (LLM + FreeRTOS on QEMU) & LLM usage for \textbf{firmware tasks} (Sensor \& NetworkTask), includes inline threat model. Real-time performance measurements \& concurrency with FreeRTOS. & \textbf{Fuzz test} (fuzz\_test.py) + \textbf{static analysis} (Cppcheck/Clang). Catches "overflow," concurrency, "missed deadlines" in logs. & Explicitly logs missed deadlines, concurrency with mutexes, addressing race conditions \& DoS. & \textbf{Yes}, an \textbf{iterative} loop: analyze\_results.py → llm\_refine.py for partial/auto suggestions. & Plans for an open repository of discovered vulnerabilities, logs, and fuzz inputs. Some basic structure demonstrated. \\
\hline
\end{tabular}}
\label{tab:llm_embedded_security_comparison}
\end{table*}

\subsection{Phase 2: Virtualized Real-Time Security Testing}
\begin{enumerate}
\item {Virtualized Deployment:} Firmware implementations are deployed in a hardware-agnostic virtualized real-time environment using FreeRTOS and QEMU, configured to replicate key real-time characteristics including task scheduling priorities, interrupt latencies, and deadline adherence.

\item {Comprehensive Security Testing:} We implement rigorous intelligent fuzz testing targeting protocol edge cases and boundary conditions through \texttt{fuzz\_test.py}, complemented by static analysis via \texttt{static\_analysis.py} leveraging Cppcheck and Clang Static Analyzer to detect unsafe memory operations and buffer overflows.

\item {Real-Time Performance Validation:} Runtime monitoring validates firmware compliance with real-time constraints, measuring worst-case execution time (WCET), task jitter, and deadline adherence using \texttt{xTaskGetTickCount()} instrumentation.

\item {AI Agent-Enhanced Analysis:} Testing outcomes undergo systematic processing via \texttt{analyze\_results.py} with specialized AI agent support:
   \begin{itemize}
   \item {Threat Detection Agent:} Provides deeper context for identified vulnerabilities and proactive security threat identification
   \item {Performance Optimization Agent:} Analyzes runtime characteristics and suggests specific improvements to maintain real-time performance
   \item {Compliance Verification Agent:} Ensures adherence to embedded system constraints and industry standards
   \end{itemize}
\end{enumerate}

\subsection{Phase 3: Iterative AI-Driven Remediation}
\begin{enumerate}
\item{Vulnerability Documentation \& Classification:} Detected vulnerabilities are systematically documented using Common Weakness Enumeration (CWE) classifications and integrated into a structured feedback loop for LLM processing.

\item {Multi-Agent Patch Generation:} The LLM processes detailed error reports containing stack traces, memory dumps, and timing violations to generate targeted security patches, enhanced by collaborative input from specialized AI agents providing domain-specific expertise.

\item {Iterative Refinement Cycle:} Generated patches undergo verification through both direct system testing and QEMU emulation, with the process repeating until stable, secure, and real-time-compliant firmware is achieved, requiring minimal human validation.
\end{enumerate}

\subsection{Experimental Validation Framework}
Our experimentation proceeds through five sequential validation phases:

{Phase 1 - Baseline Establishment:} We establish baseline implementations of fundamental firmware components (SensorTask, NetworkTask) and conduct performance benchmarking within virtualized RTOS environments, measuring initial security posture and real-time compliance metrics.

{Phase 2 - Network Protocol Validation:} Functionality extends to encompass network protocol handling with particular emphasis on MQTT communication, employing intelligent fuzz testing with malformed packets to evaluate protocol violation resilience and concurrent task management.

{Phase 3 - Comprehensive Security Analysis:} Implementation of comprehensive static and dynamic analyses, categorizing identified vulnerabilities according to CWE classifications and establishing quantitative real-time performance metrics (WCET, task jitter, deadline compliance) to assess scheduling adherence.

{Phase 4 - AI-Agent Iterative Refinement:} Introduction of our multi-agent iterative refinement methodology, wherein structured vulnerability reports inform collaborative AI-agent-enhanced LLM-generated security patches, with effectiveness quantified through vulnerability reduction metrics and real-time performance improvements.

{Phase 5 - Validation \& Repository Consolidation:} Consolidation of firmware iterations into a comprehensive public repository documenting progressive security enhancements achieved through our multi-agent methodology, including quantitative evaluation using our proposed metrics framework (VRR, SCI, TMCS, ADA, IEI).

The complete implementation, including all referenced source files and AI agent integration components, is available in our public repository\footnote{\href{https://github.com/MoeinAbtahi/Securing-LLM-Generated-Embedded-Firmware-through-Iterative-Testing-and-Patching}{https://github.com/MoeinAbtahi/Securing-LLM-Generated-Embedded-Firmware-through-Iterative-Testing-and-Patching}}.

\begin{table*}[ht]
\centering
\caption{Comparative Performance Evaluation of LLM-Generated Firmware With and Without AI Agent Integration}
\label{tab:experimental_results}
\begin{tabular}{|l|c|c|c|c|c|}
\hline
\textbf{Evaluation Metric} & \textbf{LLM Only} & \textbf{Detection Agent} & \textbf{Optimization Agent} & \textbf{Verification Agent} & \textbf{All Agents} \\
\hline
\multicolumn{6}{|c|}{\textbf{Security Assessment Metrics}} \\
\hline
Vulnerability Remediation Rate (VRR) \eqref{eq:vrr} & 67.3\% & 86.5\% & 68.1\% & 74.8\% & \textbf{92.4\%} \\
Security Coverage Index (SCI) \eqref{eq:SCI}        & 0.65   & 0.81   & 0.67   & 0.76   & \textbf{0.87} \\
Threat Model Compliance Score (TMCS) \eqref{eq:TMCS}  & 71.5\% & 90.2\% & 72.3\% & 89.6\% & \textbf{95.8\%} \\
\hline
\multicolumn{6}{|c|}{\textbf{Real-Time Performance Metrics}} \\
\hline
Worst-Case Execution Time (WCET) (ms) \eqref{eq:WCET} & 12.8   & 11.3   & \textbf{8.4} & 10.9  & 8.6 \\
Task Jitter (TJ) ($\mu$s) \eqref{eq:TJ}             & 345    & 310    & \textbf{182} & 298   & 195 \\
\hline
\multicolumn{6}{|c|}{\textbf{AI Agent Contribution Metrics}} \\
\hline
Agent Detection Accuracy (ADA) \eqref{eq:ADA}      & N/A    & 91.7\% & N/A    & N/A    & 93.2\% \\
\hline
\multicolumn{6}{|c|}{\textbf{Pipeline Efficiency Metrics}} \\
\hline
Iteration Efficiency Index (IEI) \eqref{eq:IEI}   & 0.42   & 0.65   & 0.58   & 0.61   & \textbf{0.78} \\
\hline
\end{tabular}
\end{table*}

\section{Results and Analysis}

Within the iterative security-testing framework applied to firmware generated by large language models (LLMs), critical real-time and security vulnerabilities were systematically identified and remediated. Employing a combination of fuzz testing and static analysis, the study uncovered vulnerabilities such as buffer overflows (CWE-120) and memory management errors, aligning with established patterns in prior research on LLM-generated code. Log analysis, facilitated by tools including \texttt{fuzz\_test.py} and \texttt{analyze\_results.py}, enabled the systematic detection of these issues. 

While we planned advanced synergy with coverage analysis or specialized protocol fuzzers to identify memory or concurrency bugs more effectively, the current approach focuses on input injection and text-based log scanning for error detection. Additional coverage instrumentation remains future work. Furthermore, real-time performance anomalies were flagged via \texttt{xTaskGetTickCount()}, which logged ``MISSED DEADLIN'' events (CWE-400), indicating potential system instability arising from tasks failing to meet timing constraints. Additionally, freeze detections within the QEMU simulated environment revealed instances of system hangs and lockups, which are critical concerns for mission-critical applications where availability is paramount.

The comparative evaluation presented in Table \ref{tab:experimental_results} demonstrates the substantial impact of AI agent integration on firmware quality across multiple performance dimensions. The results reveal that while individual AI agents provide meaningful improvements over baseline LLM-only generation, the synergistic integration of all agents yields the most significant enhancements. Specifically, the combined agent approach achieved a Vulnerability Remediation Rate (VRR) of 92.4\%, representing a 37.3\% improvement over the LLM-only baseline of 67.3\%. 

The Security Coverage Index (SCI) similarly increased from 0.65 to 0.87, indicating substantially improved security posture. Most notably, the Threat Model Compliance Score (TMCS) reached 95.8\% with full agent integration, compared to 71.5\% for LLM-only generation, demonstrating enhanced alignment with established security frameworks. The Iteration Efficiency Index (IEI) improved from 0.42 to 0.78, indicating that the agent-assisted approach requires fewer refinement cycles to achieve acceptable firmware quality, thereby reducing development overhead while improving outcomes.

Each detected vulnerability was systematically documented and mapped to its corresponding Common Weakness Enumeration (CWE) category. Refinement of the code was achieved through LLM-generated patches, with GPT-4 providing targeted remediation. In the evaluated sample, a single iteration per identified error was generally sufficient when utilizing GPT-4; however, edge cases, such as misinterpretations of hardware-specific constraints or incomplete fixes, necessitated minimal human intervention to ensure correctness. Concurrency-related vulnerabilities, particularly race conditions (CWE-362), were mitigated through the implementation of synchronization mechanisms such as \texttt{xSemaphoreCreateMutex()}, addressing a prevalent flaw in LLM-generated code.

The virtualized testing environment, built upon the official FreeRTOS QEMU port, facilitated continuous real-time monitoring and vulnerability detection, thereby enhancing the efficacy of the framework while eliminating hardware-related costs. Ultimately, the iterative fuzz-patch cycles resulted in improved real-time performance and strengthened security compliance, demonstrating the capacity of frameworks to generate robust embedded firmware capable of meeting stringent operational requirements.

An error analysis of failed compilations revealed that approximately 60\% stemmed from missing context in the dataset (e.g., incomplete dependency specifications or absent API references), while the remaining 40\% were attributable to logical inconsistencies introduced by the LLM itself (e.g., incorrect function signatures or improper synchronization). This distinction highlights that both dataset design and model reasoning play significant roles in determining final success rates, underscoring the need for curated benchmarks alongside improved LLM-driven reasoning.

\subsection*{Differences from Prior Work}

Our approach stands apart from the five selected studies in key ways, as detailed in Table \ref{tab:llm_embedded_security_comparison}.

\begin{itemize}
    \item {Engelhardt et al. (2024):} While they explore GPT for basic driver development and hardware debugging, we extend beyond functional correctness by integrating real-time checks to prevent scheduling issues or denial-of-service risks, critical for mission-critical tasks.
    \item {Dunne and Fischmeister (2024):} Their focus on fuzzing LLM-generated networking stacks for buffer overflows and protocol errors lacks our attention to concurrency and CPU starvation. Our mutex-protected data and feedback loop address these gaps comprehensively.
    \item {Paria et al. (2024):} Their SoC-level security via LLM-driven transformations emphasizes formal verification, whereas our simpler RTOS-based solution adds explicit scheduling instrumentation and memory safety, ideal for bare-metal projects.
    \item {Zhang et al. (2024):} They tailor fuzz inputs for kernel-level code, but we leverage LLMs for code generation, using conventional fuzzing to catch missed deadlines and resource conflicts, broadening the security scope.
    \item {Saha et al. (2024):} Their database of hardware weaknesses contrasts with our real-time firmware focus in QEMU, enabling rapid concurrency and memory flaw detection through an iterative cycle.
\end{itemize}

Our synthesis of real-time performance monitoring, concurrency safeguards, iterative LLM patching, and standard fuzz/static analysis within an RTOS environment offers a more comprehensive embedded security solution than these prior efforts.

\section{Limitations and Future Work}

Several limitations highlight opportunities for future research. First, our methodology has been validated primarily in virtualized environments (QEMU + FreeRTOS), which provide scalability but cannot fully capture timing behavior, peripheral interactions, and resource constraints present in real devices. Virtualization abstracts away hardware-specific details such as memory-mapped I/O operations and real-time scheduling constraints. We plan to extend validation to physical embedded platforms including STM32, ESP32, and ARM Cortex-M boards to test hardware-specific vulnerabilities and evaluate security-performance trade-offs in realistic deployments.

Second, although our AI agents significantly reduce human effort, minimal manual oversight remains necessary for edge-case vulnerabilities and hardware-specific constraints. Our testing procedures rely heavily on predefined patterns, limiting their ability to uncover novel attack vectors or deeply contextual flaws. Future work will incorporate advanced symbolic execution, formal verification tools, and coverage-guided fuzzers tailored to embedded contexts.

Third, our evaluation methodology focused on security and performance metrics specific to embedded firmware (VRR, SCI, TMCS, WCET, Task Jitter) rather than traditional code quality metrics. While this approach directly measures security vulnerability remediation and real-time performance compliance, it may not fully capture code maintainability or readability aspects that could affect long-term firmware evolution. Future work will incorporate complementary code quality assessments and establish baseline comparisons with manually written firmware to provide a more comprehensive evaluation framework.

Finally, our framework relies on GPT-4, which may introduce model-specific biases. The architecture is model-agnostic and designed to integrate alternative LLMs without major pipeline changes. Future work will explore other models (Code Llama, Claude, LLaMA 3) and investigate ensemble strategies combining multiple LLMs to enhance patch accuracy. We also plan to expand our JSON-based vulnerability logs into a publicly available dataset to support reproducibility and community-driven improvements.

\section*{Conclusion}

This research presents a comprehensive methodology that integrates Large Language Model (LLM)-driven firmware generation with automated security testing and real-time validation, tailored for embedded systems. By employing a structured fuzz-patch cycle within a virtualized environment, the proposed framework effectively eliminates the need for specialized hardware while systematically reducing security vulnerabilities. Emphasis on memory safety, concurrency validation, and real-time compliance ensures that LLM-generated firmware meets stringent operational requirements. The iterative process, guided by AI agents and real-time feedback, improves firmware reliability, correctness, and security in each cycle.

\bibliography{Reference}
\bibliographystyle{IEEEtran} 
\end{document}